\documentclass[prb,reprint,twocolumn,showpacs,superscriptaddress]{revtex4-1}
\usepackage{amsmath,amssymb,bm,mathrsfs,graphicx, braket, times,amsthm,enumerate}
\usepackage[colorlinks=true,citecolor=blue,linkcolor=blue]{hyperref}
\usepackage{longtable}
\usepackage{multirow}
\usepackage[all,cmtip]{xy}
\usepackage[normalem]{ulem}
\usepackage[usenames,dvipsnames]{color} 
\usepackage{multirow}

\newcommand{\sg}[1]{{\bf\em{#1}}}

\begin{document}
\title{The Lieb-Schultz-Mattis-type filling constraints in the 1651 magnetic space groups} 
\begin{abstract}
We present the first systematic study of the filling constraints to realize a `trivial' insulator symmetric under magnetic space group $\mathcal{M}$.  The filling $\nu$ must be an integer multiple of $m^{\mathcal{M}}$ to avoid spontaneous symmetry breaking or fractionalization in gapped phases. We improve the value of $m^{\mathcal{M}}$ in the literature and prove the tightness of the constraint for the majority of magnetic space groups.  The result may shed light on the material search of exotic magnets with fractionalization.
\end{abstract}

\author{Haruki Watanabe} 
\affiliation{Department of Applied Physics, University of Tokyo, Tokyo 113-8656, Japan}

\maketitle

\section{Introduction}
Putting general and non-perturbative restrictions on quantum many-body systems is valuable both as a theoretical advance and in application to material search.  The goal of this work is to further explore a general constraint on the filling $\nu$ consistent with symmetric short-range entangled (sym-SRE) phases, i.e., phases with a unique ground state with a finite excitation gap to the first excited state.  Here, the filling $\nu$ is defined as the average number of particles per unit cell, given the U(1) symmetry and the lattice translation symmetry.  The first example of theorems of this type is called the Lieb-Schultz-Mattis theorem, stating that $\nu$ must be an integer to realize sym-SRE phases~\cite{Lieb1961}.  Although the original version of the theorem was about the $S=1/2$ anti-ferromagnetic Heisenberg spin chain, it was extended to a more general class of Hamiltonians, irrespective of the form or the strength of interactions, with the particle number conservation and the translation symmetries in arbitrary spatial dimensions~\cite{Affleck1986,Affleck1988,Yamanaka,Oshikawa2000,Hastings2004,Hastings2005,PNAS,LatticeHomo}.  Among the many applications of the Lieb-Schultz-Mattis theorem, perhaps the most remarkable one is to give a coherent understanding on the Haldane conjecture~\cite{HaldanePRL,HaldanePLA,Haldane2016}: A spin-$S$ chain with vanishing magnetization may be regarded as a system of interacting particles with $\nu=S$.  When $S$ is a half integer, the system must have gapless excitations or exhibit a ground state degeneracy as a result of spontaneous symmetry breaking, while the theorem is silent for integral $S$ allowing for the Haldane phases~\cite{Affleck1986}.  In higher dimensions, the ground state degeneracy can also be accounted by a topological order, which is the hallmark of fractionalized phases~\cite{XGW}. Therefore, the Lieb-Schultz-Mattis theorem can be used as a guiding principle in search for, e.g.,  quantum spin liquids with fractional excitations~\cite{Balents,Lucile}.

Recently, the Lieb-Schultz-Mattis theorem is refined for systems with \emph{non-symmorphic} space groups (SGs)~\cite{Sid2013,PNAS}. 
It was shown that $\nu$ being an integer is, in fact, a \emph{necessary but not sufficient} condition to realize sym-SRE phases, and $\nu$ has to be an integer multiple of $m_{\text{Bbb}}^{\Gamma}=1, 2, 3, 4, \text{or }6$, depending on the specific SG of the system (Table~\ref{tab:bbb}).  In this work, we derive a filling constraint tighter than all previously known ones for systems of spinful electrons. In particular, we consider the situation where the time-reversal (TR) symmetry itself is broken, but a combination of TR and other spatial operations may be a good symmetry of the system. The results in this work are thus relevant for magnetic materials, in contrast to our previous study that assumed the direct product of an SG $\mathcal{G}$ and TR~\cite{PNAS}.  Although the absence of magnetic order is usually taken as a prerequisite for quantum spin liquids, there could be a co-existing phase of a topological order and a magnetic order.  Our result suggests that, if the magnetic order respects magnetic space group (MSG) $\mathcal{M}$ and if the filling $\nu$ does not satisfy the condition for realizing $\mathcal{M}$-symmetric sym-SRE phases, the system must be either gapless or be fractionalized.

To achieve our goal, we will rely on two related but different strategies.  The main approach defines the system of our interest on a manifold other than a torus and exposes a projective representation of a remnant symmetries on the manifold~\cite{PNAS}.  The second one utilizes two symmetry-related entanglement cuts enclosing projective representations~\cite{XieChen, PNAS}.  These ideas have been adopted only very recently~\cite{PNAS} to explore the Lieb-Schultz-Mattis-type filling constraints in systems symmetric under SG and TR separetely, and turned out to be quite successful in improving the previous results, as compared to the conventional approach of threading a magnetic flux~\cite{Oshikawa2000,Hastings2004,Hastings2005,Sid2013}. This is the first time that the new approaches are applied to MSGs.

Throughout the paper our main interest is in three spatial dimensions as it covers lower dimensional systems as well.  
We refer to specific SGs by their number assigned in the International Tables~\cite{ITC}. To distinguish SG numbers from other numerical factors, we use the \sg{bold italic} font for SG numbers.

\begin{table}[b]
\caption{\textbf{Bieberbach Bound}
\label{tab:bbb}}
\begin{tabular}{c|ccccccccccc}\hline\hline
SG $\Gamma$ & \sg{1} & \sg{4} & \sg{7} &  \sg{9} & \sg{19} & \sg{29} & \sg{33} &  \sg{76},\sg{78} &  \sg{144},\sg{145} &  \sg{169},\sg{170}\\\hline
$m_{\text{Bbb}}^{\Gamma}$ & 1& 2& 2& 2& 4 & 4& 4&4& 3& 6\\\hline\hline
\end{tabular}
\end{table}

\section{Anti-commuting symmetries on Bieberbach manifold}
Although translation-invariant systems are often discussed with the periodic boundary condition (in other words, they are defined on a torus), we can also put them on other compact flat manifolds if the symmetry of the system permits.  As shown in Ref.~\onlinecite{PNAS}, leveraging on the freedom in choice of manifolds sometimes results in tighter filling constraints.

Let us start with reviewing the derivation of $m_{\text{Bbb}}^{\Gamma}$.  Recall that the three torus can be obtained by taking the quotient of $\mathbb{R}^d$ by the translation group $T=\mathbb{Z}^d$. To get other compact flat manifolds, one can just replace the translation group with some ``fixed-point-free SG".  If none of the elements except for the identity has a fixed point, the SG is called fixed-point free. For example, an $n$-fold rotation symmetry keeps the position of the rotation axis, and a mirror reflection symmetry leaves the mirror plane unchanged.  Therefore, any SG that includes at least one of these symmetries cannot be fixed-point free.  Therefore, possible symmetry elements in fixed-point free SGs are only translations, glide reflections, and screw rotations.

\begin{figure}
\begin{center}
\includegraphics[width=0.99\columnwidth]{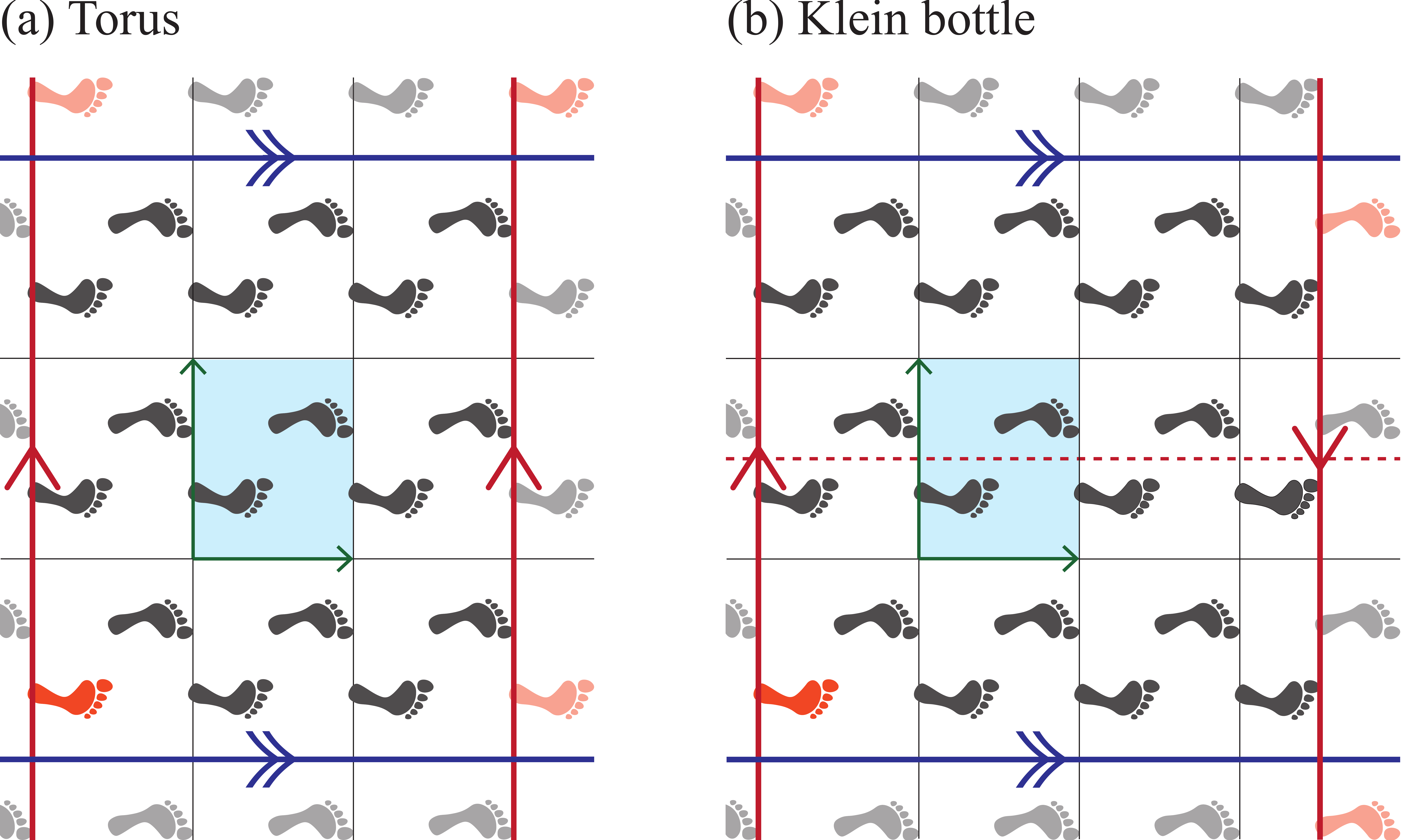}
\caption{\label{fig:kleinbottle} Boundary condition (identification rule) that leads to (a) the torus $T^2$  and (b) the Klein bottle $K$. Those colored in orange are to be identified.}
\end{center}
\end{figure}

There are only two fixed-point free SGs in 2D~\cite{ITC}. One is just the translation group and the other one has an additional glide symmetry.  Identifying the position $\vec{x}$ with $\gamma \vec{x}$ with $\gamma\in\Gamma$, these two SGs respectively produce the torus $T^2$ and the Klein bottle $K$ (see Fig.~\ref{fig:kleinbottle}).  In the same way, there are 13 different fixed-point-free SGs (including three chiral pairs)~\cite{ITC} and correspondingly there are 13 compact flat manifolds in 3D~\cite{Conway2003}.~\footnote{To see why it is important to focus on fixed-point-free SGs, let us try to identify $\tilde{\gamma} \vec{x}$  with $\vec{x}$ for every $\tilde{\gamma}\in\tilde{\Gamma}$ even when $\tilde{\Gamma}$ has a fixed-point.  In this case, the resulting set $\mathbb{R}^d/\tilde{\Gamma}$ has singularities and hence is called `orbifold' rather than a manifold.  Even when the system is in a sym-SRE phase, the properties such as the ground state degeneracy or the excitation gap may be altered by the presence of such singularities.}

The key observation is that the number of unit cells $N_\Gamma=L^d/m_{\text{Bbb}}^\Gamma$ contained in the manifold $\mathbb{R}^d/\Gamma$ can be \emph{fractional}, depending on the choice of $\Gamma$. For example, the number of unit cells on the Klein bottle illustrated in Fig.~\ref{fig:kleinbottle} (b) is $N_\Gamma=\frac{21}{2}$. However, sym-SRE phases should be insensitive to boundary conditions as long as the manifold is flat (i.e., has no curvature) and the linear dimension of the manifold $L$ is much longer than the correlation length $\xi$.  Hence, they should be well-defined on any compact flat manifold compatible with the symmetry, and in particular, the number of electrons on $\mathbb{R}^d/\Gamma$ must be integral regardless of the choice of $\Gamma$. This requirement leads to what we call the Bieberbach bound~\cite{PNAS}:
\begin{equation}
\label{eq:BbbBound}
\nu\in m_{\text{Bbb}}^\Gamma \mathbb{Z}.
\end{equation}
Namely, the filling $\nu$ must be an integer multiple of $m_{\text{Bbb}}^\Gamma$.  Every fixed-point free SG $\Gamma$ (except for the space group \sg{1} composed of the translation alone) is non-symmorphic and has $m_{\text{Bbb}}^\Gamma\geq2$ as recently studied in Refs.~\onlinecite{Sid2013,PNAS}.  For the reader's convenience, we recap the result of Ref.~\onlinecite{PNAS} in Table~\ref{tab:bbb}.

Now, suppose that the system of our interest is invariant under an SG $\mathcal{G}$ that contains a fixed-point free subgroup $\Gamma$.  
In particular, here we focus on the case where $\Gamma$ and $\mathcal{G}$ have the same translation subgroup so that they define the same unit cell. (In this case, $\Gamma$ is called a $t$-subgroup of $\mathcal{G}$.~\footnote{When $\Gamma$ is not a $t$-subgroup of $\mathcal{G}$, the unit cell of $\Gamma$ is larger than that of $\mathcal{G}$ and the filling must be converted carefully between the two groups}).  Using $\Gamma$ alone one gets the Bieberbach bound in Eq.~\eqref{eq:BbbBound}.  To achieve a tighter constraint, one should utilize elements of $\mathcal{G}$ not belonging to $\Gamma$.  To this end we demand that $\Gamma$ is a \emph{normal} subgroup of $\mathcal{G}$ so that the quotient $\mathcal{P}=\mathcal{G}/\Gamma$ forms a group.  

\begin{figure}
\begin{center}
\includegraphics[width=0.99\columnwidth]{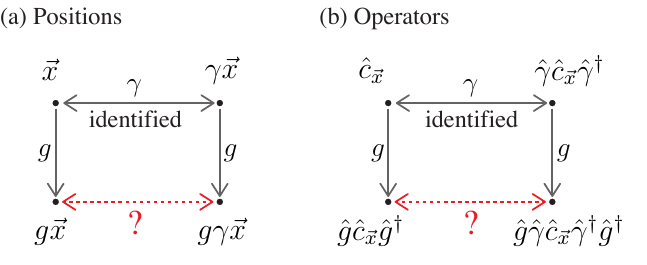}
\caption{\label{fig2} The action of $g\in\mathcal{G}$ on (a) positions and (b) operators on $\mathbb{R}^d/\Gamma$. The action is well-defined when the positions and operators in the bottom line can be identified by $\gamma'=g\gamma g^{-1}\in\Gamma$.}
\end{center}
\end{figure}

To see why it is necessary to require that $\Gamma$ is a normal subgroup of $\mathcal{G}$, recall that the manifold $\mathbb{R}^d/\Gamma$ was constructed by identifying $\gamma\vec{x}\in\mathbb{R}^d$ with $\vec{x}\in\mathbb{R}^d$ for every $\gamma\in\Gamma$. Now take an element $g\in \mathcal{G}$ that does not necessarily belong to $\Gamma$. 
Then $g$ has a well-defined action on $\mathbb{R}^d/\Gamma$ \emph{if and only if} $\gamma\vec{x}$ and $\vec{x}$ are mapped to the identical point on $\mathbb{R}^d/\Gamma$ [Fig~\ref{fig2} (a)]. In other words, there must exist an element $\gamma'\in \Gamma$ such that $g(\gamma\vec{x})=\gamma'g\vec{x}$ for all $\vec{x}$. This is equivalent with saying 
\begin{equation}
\gamma'=g\gamma g^{-1}\in\Gamma\label{position}
\end{equation}
for all $\gamma$ and $g\in\mathcal{G}$, which means that $\Gamma$ is a normal subgroup of $\mathcal{G}$.~\footnote{More generally, there can be a situation where (i) $\Gamma$ and $\tilde{\mathcal{G}}$ are $t$-subgroups of $\mathcal{G}$, (ii) $\Gamma$ is a normal subgroup of $\tilde{\mathcal{G}}$ but (iii) $\Gamma$ is \emph{not} a normal subgroup of $\mathcal{G}$. In this case, our current technique can only make use of $\tilde{\mathcal{G}}$ forgetting every elements $\mathcal{G}\setminus\tilde{\mathcal{G}}$. 
Also, there can be a few possible choices of $\Gamma$. If this is the case one should try all possible choices of fixed-point free subgroups and choose the tightest constraint. } The same consistency condition is needed at the level of operators.  Let $\hat{c}_{\vec{x}}$ be the annihilation operator of an electron at $\vec{x}$.\footnote{The notation $\hat{o}$ implies that it is an operator acting on the Hilbert space. In contrast, $g$ in $g\vec{x}$ is just a map of $\vec{x}$ to $p_g\vec{x}+\vec{t}_g$, where $p_g$ is a $d$-dimensional matrix and $\vec{t}_g\in\mathbb{R}^d$.  Finally, $\hat{g}$ for $g\in\mathcal{G}$ represents the space group operator acting on the Hilbert space that may be accompanied by U(1) phase rotation due to the projective nature of the electron spin.} When putting the system on $\mathbb{R}^d/\Gamma$ by identifying $\gamma\vec{x}$ with $\vec{x}$, one has to also identify the operator $\hat{\gamma}\hat{c}_{\vec{x}}\hat{\gamma}^\dagger$ with $\hat{c}_{\vec{x}}$ for every $\gamma\in\Gamma$. The well-defined action of $\hat{g}$ on the annihilation operators on $\mathbb{R}^d/\Gamma$ requires that 
\begin{equation}
\hat{\gamma}'=\hat{g}\hat{\gamma}\hat{g}^{-1}\text{ for }\gamma'\in\Gamma\label{operator}
\end{equation}
as illustrated in Fig~\ref{fig2} (b).

When $\mathcal{P}=\mathcal{G}/\Gamma$ satisfies both Eqs.~\eqref{position} and \eqref{operator} and thereby remains a symmetry on the manifold $\mathbb{R}^d/\Gamma$, we ask if some elements of $\mathcal{P}$ anti-commute because of the projective nature of electron spin ---  in general, anti-commuting symmetries prohibit one-dimensional representation and imply degeneracy.  As we will see below, anti-commuting symmetries in $\mathcal{P}$ improves the filling constraint by a factor of two.  Even when $\mathcal{G}$ is symmorphic, this mechanism may result in a bound tighter than the original Lieb-Schultz-Mattis theorem.

The illuminating example is SG $\mathcal{G}=\text{\sg{16}}$ ($P222$) generated by two orthogonal $\pi$ rotations about the $x$ axis $R_{2x}$ and the $y$ axis $R_{2y}$ in addition to lattice translations. Since $\mathcal{G}$ is \emph{symmorphic}, the only fixed-point free subgroup $\Gamma$ of $\mathcal{G}$ is the translation subgroup $T$ generated by $\hat{T}_x$, $\hat{T}_y$, and $\hat{T}_z$.  In this case the manifold $\mathbb{R}^3/\Gamma$ is the three torus and the Bieberbach bound is $\nu\in \mathbb{Z}$ ($m_{\text{Bbb}}^{\Gamma}=1$).  Now we examine the remnant symmetries on the torus, $\mathcal{P}=\mathcal{G}/\Gamma=222$ generated by $R_x$ and $R_y$.   For spinful electrons, $R_x$ and $R_y$ do not commute and satisfy $\hat{R}_{2x}\hat{R}_{2y}=(-1)^{\hat{N}}\hat{R}_{2y}\hat{R}_{2x}$ with $\hat{N}$ being the electron number.  Therefore, the number of electrons on the torus $N_{\rm el}=\nu N_\Gamma=\nu L^d$ must be even in order to avoid degeneracy regardless of the choice of $L$, which is possible only when $\nu\in 2\mathbb{Z}$.   This symmetry group generated by $\pi$ rotations $R_x$ and $R_y$ is known as the point group $222$.  Another example of point groups that is represented projectively by spinful electrons is $mm2$, which is generated by two orthogonal mirrors $M_x$ and $M_y$.

\begin{table}
\caption{\textbf{Bound improved by remnant projective unitary symmetries.} The asterisk ($*$) implies that U(1) symmetry is incorporated to put the system on $\mathbb{R}^3/\Gamma$.
\label{tab:bound1}}
\begin{tabular}{c|ccc}\hline\hline
SG $\Gamma$ &$\mathcal{P}=\mathcal{G}/\Gamma$ & SG $\mathcal{G}$ & $m^{\mathcal{G}}$\\\hline
 \multirow{2}{*}{\sg{1}} &$222$ & \sg{16}, \sg{21}, \sg{22}, \sg{23} & 2 \\
&$mm2$ 
& \sg{25}, \sg{35}, \sg{38}, \sg{42}  & 2\\
\hline
\multirow{1}{*}{\sg{4}} 
&$mm2$ & \sg{59}$^*$& 4\\\cline{2-4}
\hline
\multirow{2}{*}{\sg{7}} 
&$222$ & \sg{48}$^*$, \sg{50}$^*$, \sg{68}$^*$& 4\\
&$mm2$ & \sg{59}$^*$ & 4\\\cline{2-4}
\hline
\multirow{1}{*}{\sg{9}} 
&$222$ & \sg{70}$^*$ & 4\\\cline{2-4}
\hline
\multirow{1}{*}{\sg{19}} 
&$\langle B, I\rangle$ & \sg{73} & 4\\
\hline
\multirow{1}{*}{\sg{76}, \sg{78}} 
&$222$ & \sg{93}, \sg{94}, \sg{98}& 4\\
\hline
\multirow{1}{*}{\sg{169}, \sg{170}} 
&$222$ & \sg{180}, \sg{181} & 6\\
\hline\hline 
\end{tabular}
\end{table}

SG $\mathcal{G}=\text{\sg{73}}$ provides another nontrivial example of projective remnant symmetries.  This SG contains $\Gamma=\text{\sg{19}}$ as a subgroup (but \emph{not} a $t$-subgroup). Since $\mathcal{G}=\text{\sg{73}}$ contains the body-centered translation $B$, the size of the primitive unit cell of $\mathcal{G}=\text{\sg{73}}$ is a half of the unit cell of $\Gamma=\text{\sg{19}}$.  As a result, using $\Gamma=\text{\sg{19}}$ alone we can only get $m_{\text{Bbb}}^{\text{\sg{73}}}=\frac{1}{2}m_{\text{Bbb}}^{\text{\sg{19}}}=2$.~\cite{PNAS}  This bound can be improved by making use of the remnant inversion symmetry $I$.   The inversion $I$ and the body-centered translation $B$ satisfies $\hat{I}\hat{B}\hat{I}^{-1}=(-1)^{\hat{N}}\hat{B}$ on the manifold $\mathbb{R}^3/\Gamma$~\cite{PRL}.  Therefore, we get $m^{\text{\sg{73}}}=4$.

As the last example in this section, let us discuss SG $\mathcal{G}=\text{\sg{48}}$, a supergroup of SG $\text{\sg{16}}$ endowed with an additional glide symmetry $G_z$ mapping $(x,y,z)$ to $(\frac{1}{2}+x, \frac{1}{2}+y, \frac{1}{2}-z)$.  To achieve the tightest bound, one can use $\Gamma$ generated by $G_z$, $T_x$, and $T_z$ (note that $G_z^2=T_xT_y$ as elements of SG) and get a compact flat manifold $\mathbb{R}^3/\Gamma=(\text{Klein bottle})\times S^1$ with $m_{\text{Bbb}}^\Gamma=2$.  However, if we naively choose $\hat{G}_z$, $\hat{T}_x$, and $\hat{T}_z$ as the generators of $\Gamma$, $\hat{R}_{2x}$ and $\hat{R}_{2y}$ will not be remnant symmetries on $\mathbb{R}^3/\Gamma$, since
Eq.~\eqref{operator} is violated.  In fact, to respect Eq.~\eqref{operator} one has to incorporate with the U(1) symmetry and chooses $i^{\hat{N}}\hat{G}_z$, $\hat{T}_x$, $(-1)^{\hat{N}}\hat{T}_z$ as the generators of $\Gamma$.  With this choice, we can now use the projective remnant symmetry $\hat{R}_{2x}\hat{R}_{2y}=(-1)^{\hat{N}}\hat{R}_{2y}\hat{R}_{2x}$ on $\mathbb{R}^3/\Gamma$ and prove $\nu\in 2m_{\text{Bbb}}^\Gamma \mathbb{Z}$. In Table~\ref{tab:bound1}, we indicate by an asterisk ($^*$) when generators of $\Gamma$ require U(1) phases other than the fermion parity $\pm1$.

To summarize, a fixed-point free subgroup $\Gamma$ of the SG $\mathcal{G}$ allows us to define the system on a compact flat manifold $\mathbb{R}^d/\Gamma$ and thereby establishes the Bieberbach bound in Eq.~\eqref{eq:BbbBound}.  On the top of it, the remnant symmetries $\mathcal{P}=\mathcal{G}/\Gamma$, when represented projectively and anti-commutes, tighten the bound by the factor of two:
\begin{eqnarray}
&&\quad\quad \text{Fixed-point-free sub-SG } \Gamma\lhd\mathcal{G} \notag\\
&&\nearrow\notag\quad\text{Bieberbach bound}\quad \nu\in m_{\text{Bbb}}^\Gamma\mathbb{Z}.\\
\mathcal{G}\quad&&\\
&&\searrow\quad \text{Anti-commuting symmetries in } \mathcal{P}=\mathcal{G}/\Gamma\notag\\
&&\quad\quad\quad \text{Improved bound}\quad \nu\in 2m_{\text{Bbb}}^\Gamma\mathbb{Z}.\notag
\end{eqnarray}
In Table~\ref{tab:bound1}, we list 20 key SGs for which this argument enhances the bound.
The results apply to every \emph{supergroup} of these 20 SGs and tighten the bound for in total 78 out of 230 SGs in 3D.
For each SG $\mathcal{G}$, we denote by $m^\mathcal{G}\geq m_{\text{Bbb}}^\Gamma$ the best bound obtained in this way.  
To judge if the bound can be further improved or the bound is already the tightest, let $\{\nu\}_{\text{BI}}^{\mathcal{G}}$ be the set of fillings compatible with a $\mathcal{G}$-symmetric band insulator, and let $\nu_{\text{BI}}^{\mathcal{G}}$ be the greatest common divisor of $\{\nu\}_{\text{BI}}^{\mathcal{G}}$, which is worked out in Ref.~\onlinecite{PRL}.  Since a band insulator is a particular instance of sym-SRE phases, $\nu_{\text{BI}}^{\mathcal{G}}$ must be an integer multiple of $m_{\text{Bbb}}^{\Gamma}$.  In particular, we know that the bound is tight when $\nu_{\text{BI}}^{\mathcal{G}}=m^{\mathcal{G}}$ and this is indeed the case for 224 SGs out of 230.  The six exceptions are \sg{101}, \sg{102}, \sg{105}, \sg{106}, \sg{109}, and \sg{110} for which $m^{\mathcal{G}}=m_{\text{Bbb}}^{\Gamma}=2$ and $\nu_{\text{BI}}^{\mathcal{G}}=4$. For these SGs, if there exists a tighter bound or if there are intrinsically interacting sym-SRE phases is still an open question.

\begin{table}
\caption{\textbf{Bound improved by remnant anti-unitary symmetries with $\hat{\tilde{\mathcal{T}}}^2=(-1)^{\hat{N}}$.} 
\label{tab:bound2}}
\begin{tabular}{c|lcc}\hline\hline
SG $\Gamma$ &\,\,\,$\tilde{\mathcal{T}}$ & MSG $\mathcal{M}$ & $m^{\mathcal{M}}$\\\hline
 \multirow{5}{*}{\sg{1}} 
&\,\,\,$\mathcal{T}T_{\vec{a}/2}$ & \sg{1}.3 & 2\\
&\,\,\,$\mathcal{T}I$ & \sg{2}.6 & 2\\
&\,\,\,$\mathcal{T}2_1$ & \sg{4}.9 & 2\\
&\,\,\,$\mathcal{T}G$ & \sg{7}.26, \sg{9}.39 & 2\\
\hline
\multirow{6}{*}{\sg{4}} 
&\,\,\,$\mathcal{T}T_{\vec{a}/2}$ &\sg{4}.10 & 4\\
&\,\,\,$\mathcal{T}I$ & \sg{14}.78 & 4\\
&\,\,\,$\mathcal{T}2_1$ & \sg{19}.27& 4\\
&\,\,\,$\mathcal{T}G$ & \sg{29}.103, \sg{33}.148 & 4\\
&\,\,\,$\mathcal{T}4_1$ & \sg{76}.9, \sg{78}.21 & 4\\
\hline
\multirow{3}{*}{\sg{7}} 
&\,\,\,$\mathcal{T}T_{\vec{a}/2}$ & \sg{7}.27, \sg{7}.29, \sg{7}.30 & 4\\
&\,\,\,$\mathcal{T}I$ & \sg{14}.77 & 4\\
&\,\,\,$\mathcal{T}2_1$ or $\mathcal{T}G$\,\,\, & \sg{29}.101, \sg{29}.102, \sg{33}.146, \sg{33}.147 & 4\\
\hline
\multirow{1}{*}{\sg{9}} 
&\,\,\,$\mathcal{T}T_{\vec{a}/2}$ & \sg{9}.41 & 4\\
\hline
\multirow{1}{*}{\sg{19}} 
&\,\,\,$\mathcal{T}I$ & \sg{61}.437 & 8\\
\hline
\multirow{2}{*}{\sg{29}} 
&\,\,\,$\mathcal{T}T_{\vec{a}/2}$ & \sg{29}.105 & 8\\
&\,\,\,$\mathcal{T}I$ & \sg{61}.435 & 8\\
\hline
\multirow{1}{*}{\sg{33}} 
&\,\,\,$\mathcal{T}T_{\vec{a}/2}$ & \sg{33}.150 & 8\\
\hline
\multirow{1}{*}{\sg{76}, \sg{78}} 
&\,\,\,$\mathcal{T}T_{\vec{a}/2}$ &\sg{76}.11,  \sg{78}.23 & 8\\
\hline
\multirow{2}{*}{\sg{144}, \sg{145}} 
&\,\,\,$\mathcal{T}T_{\vec{a}/2}$ &\sg{144}.6, \sg{145}.9 & 6\\
&\,\,\,$\mathcal{T}6_1$ & \sg{169}.115, \sg{170}.119 & 6\\
\hline\hline 
\end{tabular}
\end{table}

\section{Black and White Magnetic space groups}
Now we turn to MSGs.  Among the in total 1651 MSGs in 3D~\cite{Bradley,BNSdatabase,Litvin,IUC}, 230 of them are identical to an SG $\mathcal{G}$ (type I MSGs), already covered in the previous section, and other 230 are simply the direct product of an SG $\mathcal{G}$ and the TR symmetry (type II MSGs or grey MSGs), whose tightest bound is studied in Ref.~\onlinecite{PNAS}.  Thus, in the following we will focus on the remaining $1651-2\times230=1191$ MSGs, called the black and white MSGs.  
The unitary part of these MSGs is identical to one of the 230 SGs and, in addition, they possess an anti-unitary element $\tilde{\mathcal{T}}=\mathcal{T}g_0$, where $g_0$ is a spatial operation that does not belong to the unitary part $\mathcal{G}$. In this paper we use the BNS notation, where an MSG is labeled by a pair of integers $\text{\sg{S}}.N$, where $\text{\sg{S}}$ refers to one of the 230 SG and $N$ is an extra label to provide the further detail.  Our study is systematic in the sense that we examine only about 40 MSGs in Tables~\ref{tab:bound2} and~\ref{tab:bound3}, rather than handling more than a thousand MSGs one by one.

The discussion follows the same logic as before.  We take a fixed-point-free sub-SG $\Gamma$ of the unitary part $\mathcal{G}$ of MSG $\mathcal{M}$ that allows us to define the system on a compact flat manifold $\mathbb{R}^d/\Gamma$. Some remnant anti-unitary symmetries in $\mathcal{P}=\mathcal{M}/\Gamma$ improve the bound by the factor of two.  For example, $\tilde{\mathcal{T}}\in\mathcal{P}$ satisfying $\hat{\tilde{\mathcal{T}}}^2=(-1)^{\hat{N}}$ or $\hat{\tilde{\mathcal{T}}}^4=(-1)^{\hat{N}}$ works, as we will see now.
\begin{eqnarray}
&&\quad\quad \text{Fixed-point-free sub-SG } \Gamma\lhd\mathcal{M}  \notag\\
&&\nearrow\notag\quad\text{Bieberbach bound}\quad \nu\in m_{\text{Bbb}}^\Gamma\mathbb{Z}.\\
\mathcal{M}\quad&&\\
&&\searrow\quad \mathcal{P}=\mathcal{M}/\Gamma \text{ with }\hat{\tilde{\mathcal{T}}}^{2n}=(-1)^{\hat{N}} \notag\\
&&\quad\quad\quad \text{Improved bound}\quad \nu\in 2m_{\text{Bbb}}^\Gamma\mathbb{Z}.\notag
\end{eqnarray}

The simplest example is given by MSG \sg{1}.3, whose unitary part $\mathcal{G}$ is just the lattice translation group.  The anti-unitary symmetry $\tilde{\mathcal{T}}=\mathcal{T}T_{\vec{a}_3/2}$ is a fractional translation by $\vec{a}_3/2$ followed by TR, satisfying $\hat{\tilde{\mathcal{T}}}^2=(-1)^{\hat{N}}\hat{T}_{\vec{a}_3}\sim(-1)^{\hat{N}}$ under a periodic boundary condition $\hat{T}_{\vec{a}_3}\sim 1$.   A slightly more nontrivial example is MSG \sg{4}.9, whose anti-unitary symmetry now comes with an additional $R_{2z}$ rotation: $\tilde{T}=\mathcal{T}T_{\vec{a}_3/2}R_{2z}$, which is a $2_1$-screw rotation followed by TR.  In this case, both $\hat{\mathcal{T}}^2$ and $(\hat{R}_{2z})^2$ separately produce a $(-1)^{\hat{N}}$ factor so that $\hat{\tilde{\mathcal{T}}}^2=(-1)^{2\hat{N}}\hat{T}_{\vec{a}_3}=\hat{T}_{\vec{a}_3}$.  Hence, the naive periodic boundary condition would imply $\hat{\tilde{\mathcal{T}}}^2\sim+1$, useless for the current purpose.   Instead, we can impose the anti-periodic boundary condition $(-1)^{\hat{N}}\hat{T}_{\vec{a}_3}\sim1$ to recover $\hat{\tilde{\mathcal{T}}}^2\sim(-1)^{\hat{N}}$.  In these examples, we get the improved bound $m^{\mathcal{M}}=2$.  We list the key MSGs in Table~\ref{tab:bound2} to which a similar argument applies.  Note that the case of MSG \sg{7}.26, generated by a glide symmetry followed by TR, has been recently discussed in Ref.~\onlinecite{YMLMO} through a flux-threading type argument.

\begin{table}
\caption{\textbf{Bound improved by remnant anti-unitary symmetries with $\hat{\tilde{\mathcal{T}}}^4=(-1)^{\hat{N}}$.} 
\label{tab:bound3}}
\begin{tabular}{c|lcc}\hline\hline
SG $\Gamma$ &\,\,\,$\tilde{\mathcal{T}}$ & MSG $\mathcal{M}$ & $m^{\mathcal{M}}$\\\hline
 \multirow{2}{*}{\sg{1}} 
&\,\,\,$\mathcal{T}R_{4z}$ &  \sg{75}.3, \sg{79}.27 & 2\\
&\,\,\,$\mathcal{T}IR_{4z}$ &  \sg{81}.35, \sg{82}.41 & 2\\
\hline
\multirow{2}{*}{\sg{7}} 
&\,\,\,$\mathcal{T}R_{4z}$ & \sg{85}.61$^*$ & 4\\
&\,\,\,$\mathcal{T}IR_{4z}$ & \sg{85}.61$^*$, \sg{86}.69$^*$ & 4\\
\hline
\multirow{1}{*}{\sg{9}} 
&\,\,\,$\mathcal{T}IR_{4z}$ & \sg{88}.83$^*$ & 4\\
\hline
\multirow{2}{*}{\sg{76}, \sg{78}} 
& $\mathcal{T}T_{(0,0,\frac{1}{2})}$ & \sg{77}.16 & 4\\
& $\mathcal{T}T_{(\frac{1}{2},\frac{1}{2},\frac{1}{2})}$ & \sg{77}.18 & 4\\
\hline\hline 
\end{tabular}
\end{table}

An anti-unitary symmetry satisfying $\hat{\tilde{\mathcal{T}}}^4=(-1)^{\hat{N}}$, such as $\mathcal{T}R_{4z}$ and $\mathcal{T}R_{4z}I$ ($R_{4z}$ is the $\frac{\pi}{2}$ rotation about the $z$ axis and $I$ is the inversion), also protects a two-fold degeneracy in a Hilbert sub-space with an odd number of electrons.   This can be seen by the fact that $\tilde{\mathcal{T}}^2$ is \emph{unitary} and that the two eigenvalues $\pm i$ of $\tilde{\mathcal{T}}^2$ are interchanged under $\tilde{\mathcal{T}}$.  More generally, an anti-unitary symmetry $\mathcal{A}$ with $\mathcal{A}^{2n}=-1$ ($n=1,2,3,\ldots$) protects two-fold degeneracy, since the eigenvalue $e^{\frac{1+2m}{n}\pi i}$ of $\mathcal{A}^2$ is paired with $e^{-\frac{1+2m}{n}\pi i}$ under $\mathcal{A}$.  Another example of $\hat{\tilde{\mathcal{T}}}^4=(-1)^{\hat{N}}$ is given by MSG \sg{77}.16 generated by the screw $S_z=T_{\vec{a}_3/2}R_{4z}$ and a half translation followed by TR $\tilde{\mathcal{T}}=\mathcal{T}T_{\vec{a}_3/2}$.  For this MSG, $\Gamma$ is generated by $\hat{S}_z$, and $\hat{\tilde{\mathcal{T}}}$ is a remnant symmetry on $\mathbb{R}^3/\Gamma$ satisfying $\hat{\tilde{T}}^4=\hat{T}_{2\vec{a}_3}=(-1)^{\hat{N}}\hat{S}_z^4\sim(-1)^{\hat{N}}$.  See Table~\ref{tab:bound3} for the list of key MSGs whose bound is improved this way.

All results in Tables \ref{tab:bound1}--\ref{tab:bound3} apply to their supergroups and we can derive our best bound for each black and white MSG $\mathcal{M}$, which we denote by $m^{\mathcal{M}}$.  We can judge if the improved bound is tight based again on band insulators.  Let $\nu_{\text{BI}}^{\mathcal{M}}$ be the greatest common divisor of the fillings consistent with an $\mathcal{M}$-symmetric band insulators, tabulated in Ref.~\onlinecite{MSGnoninteracting}.  After all of the above refinement, we find that $m^{\mathcal{M}}=\nu_{\text{BI}}^{\mathcal{M}}$ for 1010 out of 1191 black and while MSGs.  This seems to be the best we can reach in this approach.

\section{Projective representations enclosed by symmetry-related entanglement cuts}
Now let us derive a tighter bound for some of the remaining MSGs using our second method relying on the entanglement spectrum in a quasi-1D geometry.

As an illustrative example, let us discuss MSG \sg{25}.61 generated by two orthogonal mirrors $M_x$ and $M_y$ (which flip $x$ and $y$, respectively) and a half-lattice translation in $z$ followed by TR $\tilde{\mathcal{T}}=\mathcal{T}T_{(0,0,\frac{1}{2})}$.  We take the periodic boundary condition in $x$ and $y$ but leave $z$ infinitely long.  On the one hand, if the ground state is in a sym-SRE phase, the entanglement spectrum at the cut $z=z_0$ (i.e., tracing-out everything in, say $z>z_0$) must have a discrete Schmidt weight. The Schmidt weight at the cut $z=z_0$ and $z=z_0+\frac{1}{2}$ must be identical because of the (anti-unitary) half-translation symmetry $\tilde{\mathcal{T}}$. On the other hand, the two mirrors anti-commute as explained above, and when the number of electrons between the two entanglement cuts at $z=z_0$ and $z=z_0+\frac{1}{2}$ is odd, the Schmidt weight fails to be (half-)translation symmetric as the projective representation alters a `doublet' into `singlets' and vice versa~\cite{PNAS, XieChen}.  Therefore, to realize sym-SRE phases, there needs to be an even number of electrons between the two cuts irrespective of the period in $x$ and $y$, which is only possible when the filling $\nu\in4\mathbb{Z}$.  

More generally, we can improve the Bieberbach bound by a factor of $2$ when two symmetry-related entanglement cuts enclose a projective representation of symmetries that leave the two entanglement cuts unmoved.  This argument proves the tightest bound for about 70 MSGs listed in Appendix~\ref{app:1}.

Let us now clarify the relation between our first and the second approaches to tightening the filling constraints.  When the two entanglement cuts in the second approach are related by a unitary symmetry $g$, the same bound can be obtained by the first approach by putting the system on the manifold $\mathbb{R}^d/\Gamma$ where $\Gamma$ is generated by $g$ and translations.  However, when the two cuts are related by an anti-unitary symmetry $a$, to our knowledge there is no well-defined way to introduce a boundary condition using $a$.  This is why we could achieve a tighter bound in the second approach in the above examples.  However, it is not clear how one can make use of multiple nonsymmorphic symmetries simultaneously in the second approach. As a result, the second approach is weaker than the first one when it is mandatory to combine of multiple nonsymmorphic elements to derive a nontrivial bound.  At this moment, the first and second approaches give the same constraints in some cases and are complementary to each other in other cases. Exploring their relations in more details is certainly an interesting future problem.

\section{Conclusion}
In this paper, we explored the Lieb-Schultz-Mattis-type filling constraints in systems symmetric under MSGs, having in mind the application of searching magnetic materials realizing symmetry-protected topological phases and/or (symmetry-enriched) topological orders.  We systematically derived a filling constraint tighter than all previously known ones, and proved the saturation $m^{\mathcal{M}}=\nu_{\text{BI}}^{\mathcal{M}}$ (i.e., no further improvement is possible) relying on $\mathcal{M}$-symmetric band insulators~\cite{MSGnoninteracting} for the majority of MSGs.    However, there are still 118 type III or IV MSGs among 1191 for which we could not prove the tightness of the bound. We list them in Appendix~\ref{app:2}. In particular, MSG \sg{49}.271 (type III) and \sg{51}.293 (type III) can be regarded as a symmetry of electrons on a 1D lattice but with the standard 3D spin. Thus one might be able to analyze an electronic system symmetric under one of these symmetries by bosonization, and answer a fundamental problem of whether there can possibly exist an intrinsically-interacting sym-SRE phases that can never be realized by band insulators.  We leave this exciting open question for future work.

\begin{acknowledgments}
The author thanks H. C. Po for useful discussions on this topic. This work was supported in part by JSPS KAKENHI Grant Numbers JP17K17678.
\end{acknowledgments}

\bibliography{references}

\appendix

\section{MSGs covered by entanglement arguments}
\label{app:1}
Here we list those MSGs for which the filling constraint is improved by the entanglement argument. We find that two projective symmetries work. The first one is $mm2$, and the other one is the point group generated by $\mathcal{T}R_{4z}$. These groups have in common one property that they leave the plane for the entanglement cut invariant.
\subsection{Entanglement argument of $mm2$ type}
\sg{25}.61, \sg{25}.64, \sg{25}.65, \sg{35}.169, \sg{35}.171, 
\sg{47}.254, \sg{47}.255, \sg{47}.256, \sg{65}.488, \sg{65}.490, 
\sg{99}.168, \sg{99}.170, \sg{101}.179, \sg{101}.181, \sg{102}.187,  \sg{102}.189, \sg{105}.211, 
\sg{105}.214, \sg{106}.219, \sg{106}.222, \sg{109}.239, \sg{109}.242, \sg{110}.245, \sg{110}.248, 
\sg{111}.256, \sg{111}.258, \sg{113}.274, \sg{115}.288, 
\sg{115}.290, \sg{123}.348, \sg{123}.349, \sg{123}.350, 
\sg{131}.437, \sg{131}.439, \sg{131}.442, \sg{132}.449, 
\sg{132}.450, \sg{132}.452, \sg{133}.461, \sg{133}.466, 
\sg{134}.473, \sg{134}.476, \sg{135}.487, \sg{136}.498, 
\sg{137}.509, \sg{137}.514, \sg{141}.553, \sg{141}.558, 
\sg{183}.190, \sg{191}.242, \sg{200}.17, \sg{215}.73, \sg{221}.97, \sg{223}.107, \sg{224}.112, \sg{227}.130.

\subsection{Entanglement argument of $\mathcal{T}R_{4z}$ type}
\sg{103}.197, \sg{103}.198, \sg{104}.205, \sg{104}.206, \sg{124}.354, \sg{124}.355, \sg{124}.356, \sg{124}.358, \sg{126}.380, \sg{126}.382, \sg{128}.402, \sg{128}.403, \sg{222}.100.

\section{Remaining type III or IV MSGs}
\label{app:2}
The followings are the list of 118 type III or IV MSGs for which we could not establish $m^{\mathcal{M}}=\nu_{\text{BI}}^{\mathcal{M}}$.
\subsection{$m^{\mathcal{M}}=2$ and $\nu_{\text{BI}}^{\mathcal{M}}=4$ (89 MSGs)}
\sg{16}.5, \sg{16}.6, \sg{25}.63, \sg{38}.194, \sg{44}.233, \sg{48}.261, \sg{49}.271, \sg{50}.283, \sg{51}.293, \sg{59}.408, \sg{63}.460, \sg{66}.497, \sg{68}.517, \sg{70}.531, \sg{71}.538, \sg{72}.545, \sg{73}.551, \sg{74}.557, \sg{84}.54, \sg{86}.70, \sg{88}.84, \sg{89}.93, \sg{89}.94, \sg{90}.99, \sg{94}.131, \sg{98}.161, \sg{99}.169, \sg{100}.173, \sg{106}.221, \sg{106}.223, \sg{107}.232, \sg{108}.235, \sg{110}.247, \sg{110}.249, \sg{111}.257, \sg{112}.262, \sg{112}.266, \sg{113}.269, \sg{114}.277, \sg{114}.278, \sg{115}.289, \sg{116}.293, \sg{117}.301, \sg{118}.309, \sg{119}.320, \sg{120}.323, \sg{122}.335, \sg{122}.336, \sg{124}.359, \sg{125}.368, \sg{125}.371, \sg{126}.383, \sg{127}.390, \sg{129}.413, \sg{129}.418, \sg{131}.443, \sg{132}.455, \sg{133}.464, \sg{133}.465, \sg{134}.479, \sg{135}.486, \sg{135}.489, \sg{137}.515, \sg{139}.540, \sg{140}.544, \sg{140}.546, \sg{140}.549, \sg{141}.559, \sg{142}.567, \sg{192}.251, \sg{193}.256, \sg{194}.267, \sg{195}.3, \sg{201}.20, \sg{203}.28, \sg{207}.43, \sg{210}.54, \sg{216}.77, \sg{218}.83, \sg{218}.84, \sg{219}.87, \sg{220}.91, \sg{222}.102, \sg{223}.108, \sg{224}.114, \sg{225}.121, \sg{226}.125, \sg{226}.126, \sg{227}.132.

\subsection{$m^{\mathcal{M}}=4$ and $\nu_{\text{BI}}^{\mathcal{M}}=8$ (28 MSGs)}
\sg{50}.286, \sg{50}.288, \sg{59}.413, \sg{59}.414, 
\sg{101}.186, \sg{102}.192, \sg{105}.217, \sg{109}.244, 
\sg{125}.374, \sg{126}.385, \sg{129}.420, \sg{130}.426, 
\sg{131}.445, \sg{132}.458, \sg{133}.470, \sg{134}.480, 
\sg{135}.491, \sg{136}.504, \sg{137}.516, \sg{137}.517, 
\sg{138}.530, \sg{141}.560, \sg{142}.564, \sg{142}.569, 
\sg{227}.133, \sg{228}.137, \sg{228}.138, \sg{230}.149

\subsection{$m^{\mathcal{M}}=2$ and $\nu_{\text{BI}}^{\mathcal{M}}=8$ (only 1 MSG)}
\sg{133}.467

\clearpage

\end{document}